\documentstyle[preprint,pre,aps]{revtex}
\begin{document}
\draft
\title
{Exact solutions of a restricted ballistic deposition model
                     on a one-dimensional staircase }
\author{Hyunggyu Park}
\address{Department of Physics, Inha University,
Inchon 402-751, Korea}
\author{Meesoon Ha and In-mook Kim}
\address{Department of Physics, Korea University,
Seoul 136-701, Korea}
\maketitle
\begin{abstract}
Surface structure of a restricted ballistic deposition(RBD) model
is examined on a one-dimensional staircase
with free boundary conditions. In this model,
particles can be deposited only at the steps of the staircase.
We set up recurrence relations
for the surface fluctuation width $W$ using generating function
method.
Steady-state solutions are obtained exactly given
system size $L$. In the infinite-size limit,
$W$ diverges as $L^\alpha$ with the scaling exponent
$\alpha=\frac{1}{2}$.
The dynamic exponent $\beta$ $(W\sim t^\beta)$ is also found to be
$\frac{1}{2}$ by solving the recurrence relations numerically.
This model can be viewed as
a simple variant of the  model
which belongs to the Kardar-Parisi-Zhang (KPZ) universality class
$(\alpha_{KPZ}= \frac{1}{2} , \beta_{KPZ}=\frac{1}{3})$.
Comparing its deposition time scale with that of the single-step
model,
we argue that $\beta$ must be the same as
$\beta_{KPZ}/(1-\beta_{KPZ})$,
which is consistent with our finding.
\end{abstract}
\section{Introduction}

Recently, dynamics of growing surfaces or interfaces has been
one of the most interesting
problems in surface science. Crystal growth, dielectric
breakdown, fluid displacement in porous media, vapor deposition,
spray
painting and coating, biological growth and electrodeposition
are only a few examples which are important from
both theoretical and practical point of view \cite{G,DF}.
Stochastically growing surface exhibits
nontrivial scaling behavior and evolves to
a steady state without
any characteristic time or spatial scale. This observation has led to
the development of general scaling approach for describing growing
surface
which exhibits a self-affine fractal geometry \cite{M}. In particular,
the dynamic scaling approach \cite{F,BD} has been applied to the
study of
a variety of theoretical models of growing surfaces.

The important feature of growing surface is
the roughness of surface induced by stochastic noises and
its roughness can be characterized by few scaling exponents.
A fundamental question is how to classify the rough surfaces
given dynamic rules of deposition processes.
The simplest model is the random deposition model which has no
interactions between neighboring columns. This model can be
solved exactly for any system size.
It produces a rough
surface in which the surface width grows with the square root of time
but never reaches a steady state for finite systems.
When interactions are introduced between columns, the surface width
for finite systems saturates in the long-time limit. The evolution of
the surface width $W$ is described by a dynamic scaling relation
\begin{equation}
\label{eq-1}
W = L^{\alpha }f(t/L^z),
\end{equation}
where $L$ is the linear dimension of the system, $t$ is the
time, and the scaling function $f$ has the asymptotic behaviors
$f(x)\sim
x^{\beta}$, with $\beta =\alpha /z$, for $x\ll 1$, and $f(x)\sim $
constant for $x \gg 1$.  Consequently the limiting behavior of the
width are
\begin{eqnarray}
\label{eq-2}
  W\sim \cases{L^{\alpha}&$(t\gg L^z)$\cr
                t^{\beta}&$(t\ll L^z)$\cr}.
\end{eqnarray}

Many stochastic models with intercolumnar interactions have
been introduced and extensively studied.
One simple example is the random deposition model with
surface diffusion \cite{F'}. A deposited particle
diffuses on the surface and is incorporated into the deposit
upon reaching a local minimum. This rearrangement
process reduces the surface
tension of the deposit and the surface becomes
relatively smooth with scaling exponents $\alpha\simeq \frac{1}{2}$
and $\beta\simeq \frac{1}{4}$ in $(1+1)$ dimensions. These values
are obtained by numerical simulations and the exact solutions are
not available as yet. It has been argued
that this model can be described
by a continuum equation with a noise source
\begin{equation}
\label{eq-3}
{\partial h\over {\partial t}}=\nu \nabla ^2h+
\eta,
\end{equation}
where $h$ denotes the deviation of the local height of the deposit
from
its average value and $\eta$ is a Gaussian random noise.
The solution of eq.(\ref{eq-3}) yields
$\alpha =1/2$ and $\beta =1/4$ in
$(1+1)$ dimensions \cite{EW}, which is consistent with numerical
results.
Many other models with different rearrangement rules have been
studied
by numerical simulations and also by solving corresponding
continuum equations. However, both methods have their own weak points.
Numerical simulations often do not reach the
scaling regime within a reasonable amount of time.
A continuum equation corresponding to a given stochastic model
may be reasonably guessed but it is very difficult
to prove that it correctly describes the stochastic model.
For example,
there has been a great deal of controversies over which
continuum equation describes
the Wolf-Villain model \cite{WV}. It is important to find exact
solutions
for stochastic models but it is usually formidable.

Some other stochastic models like the ballistic deposition model
\cite{BD,BD'},
the single-step model \cite{BD,CW,PR},
and the restricted solid-on-solid model \cite{RSOS}
behave differently from the particle-rearrangement models.
Numerical simulations on these models
provide the scaling exponents $\alpha\simeq \frac{1}{2}$
and $\beta\simeq \frac{1}{3}$ in $(1+1)$ dimensions.
In these
models, the lateral current of particles along the surface
is not conserved, so evolution of the surface involves
the lateral growth. It has been argued
that all these models can be described
by the Kardar-Parisi-Zhang (KPZ) equation \cite{KPZ}
\begin{equation}
\label{eq-4}
{\partial h\over {\partial t}}=\nu \nabla ^2h+
  \frac{\lambda}{2} (\nabla h)^2 + \eta
\end{equation}
where the nonlinear term represents the lateral growth.
Dynamic renormalization group calculations \cite{RG} for
eq.(\ref{eq-4}) yield
$\alpha =1/2$ and $\beta =1/3$ in
$(1+1)$ dimensions (KPZ universality class),
which is consistent with numerical results.
However, higher-dimensional cases are not settled down yet.

In this paper, we present a nontrivial
stochastic model which can be solved exactly for finite system sizes
in $(1+1)$ dimensions. This model is similar to the ballistic
deposition model, but height differences between neighboring
columns are restricted to be positive or zero:
$h_{i+1}-h_i\ge 0$ where $h_i$ is the column height at site $i$.
The resulting surface should look like a one-dimensional
ascending staircase with variable widths of terraces and
variable heights of steps (see Fig.1).
Particles can be deposited only at the steps of the
staircase, otherwise dynamic processes are rejected.
{}From the master equation of this model with
open boundary conditions, we set up recurrence relations
for the surface fluctuation width $W$ using generating function
method.
In the steady-state limit, we are able to extract the exact solutions
for system size $L$ from which we
obtain $\alpha=\frac{1}{2}$ in the infinite-size limit. We also find
$\beta=\frac{1}{2}$ by solving the recurrence relations numerically.
No statistical errors are involved in this estimate because
the ensemble average has already been done in the recurrence
relations.
It is somewhat surprising at first glance because
this model can be viewed as a simple variant of the single-step model
(KPZ universality class) \cite{BD,PR}
by rotating the surface clockwise by 45$^\circ$
but the values of the exponents are different from those of
the single-step model.
We argue that the dynamic exponent
$\beta$ must be the same as $\beta_{KPZ}/(1-\beta_{KPZ})$
by comparing its deposition time scale with that of the single-step
model,
which is consistent with our finding. Considering this
argument other way around, one can say that our results for
these scaling exponent are
the best estimates for the stochastic models in the
KPZ universality class.

The outline of this paper is as follows. In section II, we describe
our model
and discuss its relation to a particle dynamics model,
a spin-exchange model and the single-step model.
In section III, we define the generating function and set up
the recurrence relations. In section IV, we find the exact solutions
in the steady-state limit and present the numerical estimate for
$\beta$.
Comparison with the single-step model is given in section V and
we conclude in section VI with a brief summary.

\section{Restricted ballistic deposition model}

Consider the ballistic deposition (BD) model \cite{BD} in $(1+1)$
dimensions.
In the  BD model, a particle falls down along a straight line
and sticks to the surface of the deposit.
We introduce a restricted
ballistic deposition model (RBD) by allowing only deposition
processes which preserve the morphology of an ascending staircase
of the surface, i.e.
\begin{equation}
\label{eq-5}
n_i(t)\equiv h_{i+1}(t)-h_i(t) \ge 0 \qquad\hbox{for all}\quad i,
\end{equation}
where $h_i(t)$ is the column height of site $i$ at time $t$ and
$n_i(t)$ is the height of the step between $i$-th and $(i+1)$-th
sites.

We start with a surface of horizontal length $L$ with unit step
heights
at every possible steps, i.e.~ $n_i(0)=1$ for $i=1,2,\cdots,L-1$,
equivalently $h_i(0)=i$ for $i=1,\cdots,L$.
(The reason why we take this particular configuration as an initial
state will be discussed in section IV).
First, select a site $i$
randomly and drop a particle along a straight line.
The particle becomes part of the deposit when
it contacts a particle in the deposit, only if the resulting
step heights are all positive or zero (eq.(\ref{eq-5})).
Otherwise, the
particle is rejected(see Fig.1).
When a site $i(\neq 1, L)$ is chosen at time $t$,
the deposition process can occur for $n_i(t) > 0$
and the rejection process for $n_i(t)= 0$.
Then $n_i$ and $n_{i-1}$ are updated as
\begin{eqnarray}
\label{eq-6}
  n_i(t+\Delta t) &= &0,  \nonumber\\
 n_{i-1}(t+\Delta t) &=&n_{i-1}(t)+n_i(t),
\end{eqnarray}
where $\Delta t$ is the time elapsed during one deposition attempt.
Notice that eq.(\ref{eq-6}) is valid for both deposition and
rejection processes. This is the key point of obtaining
rather simple recurrence relations to be solved exactly.

We use {\em free} boundary conditions. When a site $i=1$ is chosen,
the deposition and rejection process can be described by
the first equation of eq.(\ref{eq-6}). At site $L$, only
the deposition process can occur (no rejection) such that
\begin{equation}
\label{eq-7}
n_{L-1}(t+\Delta t) = n_{L-1} + 1.
\end{equation}
This equation can be identified as the second equation of
eq.(\ref{eq-6})
by defining $n_L\equiv 1$. (This is useful to write simple
recurrence relations in section III).

The RBD model can be mapped on a one-dimensional particle
dynamics model with directed diffusion and mass-conserving
coalescence processes. Dynamic variable $n_i$ represents
the mass of the particle at site $i$. Eq.(\ref{eq-6})
implies that a particle of mass $n_i(t)$ at site $i$ diffuses to
the left
and stays there if $n_{i-1}(t)=0$ (no particle at site $i-1$
at time $t$)
or coalesces with a particle of
mass $n_{i-1}(t)$ at site $i-1$ if $n_{i-1}(t) > 0$.
At the boundaries, there are input (source) and output
(sink) of particles.
Particles of unit mass come into the system from
the right side of site $L-1$ and
particles of arbitrary mass leave the system to
the left side of site 1 as shown in Fig.1.

One can map this RBD model onto a variant of  spin-exchange models,
in the
same way as the Toom model \cite{TOOM1,TOOM2} in the low-noise limit
is mapped onto one of the spin-exchange models.
Details of this mapping can be found in reference (16).
Our model does not belong to general
${\cal M}^{(k)}$ spin-exchange models(see Appendix A of \cite{TOOM2}).
This aspect will
be discussed further elsewhere.

One can find close resemblance between the RBD model and
the single-step model if not {\em sticky} particles but {\em smooth}
particles
are used in the RBD model. Now a particle falls down along a
straight line and becomes part of the deposit when it hits
the ground.
Of course, the surface of the RBD model
is rotated by 45$^\circ$ from the surface of the single-step model.
Both models restrict depositions only at the steps (or valleys
on the 45$^\circ$-rotated surface, see Fig.2).
Unfortunately our model with smooth particles does not yield
simple recurrence relations in contrast to the RBD model with
sticky particles. Dynamics with smooth particles can not be
represented by simple equations like eq.(\ref{eq-6}) but
equations with conditional statements.
However, we argue in section V that
these two models exhibit essentially the same scaling behavior
if time-scale difference is taken into account.

\section{Generating function and recurrence relations}

In this section, we write the master equation for the evolution
of the surface in the RBD model with general
{\em inhomogeneous} deposition rates.
The deposition rates are parameterized by $\{\kappa_1,\kappa_2,\cdots,
\kappa_L\}$ where $\kappa_i$ is a relative deposition rate at site $i$
with normalization $\sum_i \kappa_i = 1$. The RBD model with
ordinary homogeneous deposition rates is restored when all
$\kappa_i$'s
are replaced by $1/L$.
A surface configuration can be characterized by a
state vector ${\bf n}=(n_1,n_2,\cdots,n_{L-1})$ if
we are not concerned
about the overall average height but are interested in
relative heights only.
We define $P_t ({\bf n})$ as a probability to find a configuration
${\bf n}$ at time $t$.
Then the master equation
can be written as
\begin{eqnarray}
\label{eq-8}
P_{t+{\Delta t}} ({\bf n}) &=& \kappa_1 \delta_{n_1,0}
          \sum_{m=1}^{\infty} P_t (m,n_2,\cdots)
     + \kappa_L \theta (n_{L-1}) P_t (n_1,\cdots,n_{L-1}-1)
\nonumber\\
&+& \sum_{i=2}^{L-1} \kappa_i \theta(n_{i-1})\delta_{n_i,0}
\sum_{m=1}^{n_{i-1}} P_t (\cdots,n_{i-2}, n_{i-1}-m,m,n_{i+1},\cdots)
    \nonumber\\
     &+& \sum_{i=1}^{L-1} \kappa_i \delta_{n_i,0} P_t ({\bf n}),
\end{eqnarray}
where $\delta$ is a Kronecker delta function and $\theta(x)=1$ for
$x\ge 1$ and zero otherwise. The first three terms in the right
hand side of the above equation describe the deposition processes
at the boundaries and inside the boundaries. The last term describes
the rejection processes.

Consider the generating function defined as
\begin{equation}
\label{eq-9}
  G_t(z_1,z_2,\cdots,z_{L-1})\equiv\sum_{n_i\ge 0}
                        z_1^{n_1}z_2^{n_2}\cdots z_{L-1}^{n_{L-1}}
                        P_t({\bf n}).
\end{equation}
Using eq.(\ref{eq-8}), one can obtain the generating function
at time $t+{\Delta t}$ in terms of $G_t$
\begin{eqnarray}
\label{eq-10}
  G_{t+{\Delta t}}(z_1,\cdots,z_{L-1})& = &\kappa_1 G_t(1,z_2,\cdots)
                   + \kappa_L z_{L-1}G_t(z_1,\cdots,z_{L-1})
\nonumber\\
                  & + &\sum_{i=2}^{L-1}\kappa_i
                    G_i(\cdots,z_{i-1},z_{i-1},z_{i+1},\cdots).
\end{eqnarray}
The ensemble averages of step heights (particle masses)
and two-point(mass-mass) correlation functions are obtained
by differentiating the generating function and setting all
$z_i$'s to be 1 as
\begin{eqnarray}
\label{eq-11}
   \langle n_i\rangle_t &=&
\frac{\partial G_t}{\partial z_i}\bigg|_{\{ z \}=1},
\nonumber\\
   \langle n_in_j \rangle_t&=&\frac{\partial}{\partial z_i}
           \biggl(z_j \frac{\partial G_t}{\partial z_j}\biggr)\bigg|_
                  {\{ z \}=1}.
\end{eqnarray}
Thus the dynamics of the step heights (masses) and
two-points correlation functions are given as
\begin{eqnarray}
\label{eq-12}
  \langle n_i\rangle_{t+\Delta t}& = &(1-\kappa_i)
\langle n_i\rangle_t
+ \kappa_{i+1}\langle n_{i+1}\rangle_t,
\nonumber\\
  \langle n_i^2\rangle_{t+\Delta t}& = &
(1-\kappa_i)\langle n_i^2\rangle_t
              + 2\kappa_{i+1}\langle n_in_{i+1}\rangle_t
                 +\kappa_{i+1}\langle n_{i+1}^2\rangle_t,
\nonumber\\
  \langle n_in_{i+1}\rangle_{t+\Delta t}& = &
(1-\kappa_i-\kappa_{i+1})
                      \langle n_in_{i+1}\rangle_t
                           +\kappa_{i+2}\langle n_in_{i+2}\rangle_t,
\\
  \langle n_in_j\rangle_{t+\Delta t}& = &
(1-\kappa_i-\kappa_j)\langle n_in_j\rangle_t
           + \kappa_{i+1}\langle n_{j+1}n_j\rangle_t
                +\kappa_{j+1}\langle n_in_{j+1}\rangle_t.
\nonumber
\end{eqnarray}
where $i,j=1,\cdots,L-1$ with $|i-j|\ge 2$ and $n_L$ is set to be 1.

First, we consider the steady-state only. The ensemble-averaged masses
in the steady state, $\langle n_i\rangle_\infty$, satisfy the
recurrence relations
\begin{equation}
\label{eq-13}
  \kappa_i\langle n_i\rangle_{\infty}
=\kappa_{i+1}\langle n_{i+1}\rangle_{\infty}
           =\cdots=\kappa_L\langle n_L\rangle_{\infty}=\kappa_L,
\end{equation}
which leads to
\begin{equation}
\label{eq-14}
  \langle n_i\rangle_{\infty} = \frac{\kappa_L}{\kappa_i}.
\end{equation}
In the homogeneous case,
$\langle n_i\rangle_{\infty}=1$ at any site $i$, so
the ensemble-averaged
surface looks like an ascending {\em regular} staircase with
unit step heights and unit terrace widths
in the steady-state (long-time) limit.

We also find recurrence relations for two-point correlation functions
in the steady state as
\begin{eqnarray}
\label{eq-15}
  Q_{iL}& = &Q_{i+1,L},
\nonumber\\
  Q_{ii}& = &w_iQ_{i,i+1}+Q_{i+1,i+1},
\nonumber\\
  Q_{i,i+1}& = &u_{i,i+1}Q_{i,i+2},
\\
  Q_{ij}& = &u_{ij}Q_{i+1,j}+v_{ij}Q_{i,j+1},
\nonumber
\end{eqnarray}
where $Q_{ij}$, $u_{ij}$, $v_{ij}$, and $w_i$ are defined as
\begin{equation}
\label{eq-16}
  Q_{ij} \equiv \kappa_i \langle n_i n_j\rangle_{\infty},
\end{equation}
\begin{equation}
\label{eq-17}
  u_{ij} \equiv \frac{1}{1+ \kappa_j/\kappa_i},\qquad
  v_{ij} \equiv u_{ij}\frac{\kappa_{j+1}}{\kappa_i},
\qquad\mbox{\rm and}
      \qquad w_i \equiv 2 \frac{\kappa_{i+1}}{\kappa_i},
\end{equation}
with $Q_{ji}=\frac{\kappa_j}{\kappa_i} Q_{ij}$.
It is easy to solve the above recurrence relations
numerically for finite values of $L$.
Start from the known value of $Q_{LL}=\kappa_L$.
Using the recurrence relations repeatedly, one can find the values of
$Q_{L-1,L}$, then $Q_{L-2,L}$ and $Q_{L-1,L-1}$, then
$Q_{L-3,L}$ and $Q_{L-2,L-1}$, then so on. $Q_{ij}$ is determined
by $Q_{i^\prime j^\prime}$'s only for $i^\prime+j^\prime > i+j$.
This brings out an idea of mapping these recurrence relations
to a directed random walk problem.

Consider an upper diagonal half of a $L\times L$ square lattice
(Fig.3).
We assign appropriate weights; $u_{ij}$, $v_{ij}$, $w_i$, 1, or 0
on the bonds connecting neighboring sites. Now imagine that a walker
starts from a lattice site $(L,L)$ and moves only left or down
(directed random walk).
There are numerous paths along which the walker can reach a site
$(i,j)$. A typical path is drawn in Fig.3. Paths going through
a lower diagonal half of the square lattice are excluded.
The weight to a path
is given as the product of bond weights along the path. Then,
$Q_{ij}$ can be calculated by summing up path weights for all
possible paths. So the formal solutions for $Q_{ij}$ are given as
\begin{equation}
\label{eq-18}
  Q_{ij}=Q_{LL} \sum_{\hbox{\scriptsize all}
                        \atop{\hbox {\scriptsize possible paths}}}
                \prod_{\hbox{\scriptsize a path}}
                (u_{ij},v_{ij},w_i,1,0).
\end{equation}
This is similar to the first passage problem \cite{Feller}
when all bond weights are set to be equal.
We are able to extract exact solutions in the case of homogeneous
deposition rates (i.e.~$\kappa_i=1/L$  for all site $i$).

\section{Exact solutions and critical exponents}

We consider the homogeneous case only. Then $u_{ij}\equiv u=1/2$,
$v_{ij}\equiv v=1/2$, and $w_i\equiv w=2$. And $Q_{ij}=Q_{ji}$.
For convenience, we define normalized two-point correlation functions
${\tilde Q}_{ij}\equiv Q_{ij}/Q_{LL}$ where
$Q_{LL}=1/L$.
{}From eq.(\ref{eq-18}), ${\tilde Q}_{ij}$ is the sum of
path weights over
all possible directed paths from $(L,L)$ to $(i,j)$ which do not
go through a lower diagonal half of the $L\times L$ square lattice.
Some of these paths have the same path weights but some do not.
So it is necessary to distinguish the paths with different path
weights.

First, consider the case $i<j$ $(j\le L-1)$.
Paths going through a bond connecting the points $(\ell, L)$ and
$(\ell, L-1)$ ($\ell\ge i$) must have the same path weights,
\begin{equation}
\label{eq-19}
  {\cal W} (i,j;\ell)=
1^{L-\ell} u^{L-j} v^{\ell-i}=\left(\frac{1}{2}\right)^{L+\ell-i-j}.
\end{equation}
The number of such paths,
${\cal N}(i,j;\ell)$, is
equivalent to the number of directed paths from $(\ell, L-1)$ and
to $(i,j)$ which do not touch the diagonal line. Paths touching the
diagonal line do not contribute to ${\tilde Q}_{ij}$ for $i<j$. Using
the reflection principle \cite{Feller}, the number of paths can be
easily obtained
as
\begin{equation}
\label{eq-20}
 {\cal N} (i,j;\ell)=
\left( \begin{array}{c} \ell-i+L-1-j \\ L-1-j \end{array} \right)
-\left( \begin{array}{c} \ell-i+L-1-j \\ L-1-i \end{array} \right),
\end{equation}
where $\left(\cdots\right)$ is a combinatorial factor.
Introducing new site variables for convenience
($p\equiv L-i-1$, $q\equiv L-j-1$, $r\equiv L-\ell-1$),
the normalized correlation functions are
\begin{eqnarray}
\label{eq-21}
{\tilde Q}_{ij} &=&
\sum_{\ell=i}^{L-2} {\cal W} (i,j;\ell) {\cal N} (i,j;\ell) \\
  &=& \frac{1}{2} \sum_{r=1}^{p}
  \left( \begin{array}{c} p+q-r \\ q \end{array} \right)
  \left(\frac{1}{2}\right)^{p+q-r}
- \frac{1}{2}\sum_{r=1}^{q}
  \left( \begin{array}{c} p+q-r \\ p \end{array} \right)
  \left(\frac{1}{2}\right)^{p+q-r}. \nonumber
\end{eqnarray}
This equation  can be rewritten as
\begin{equation}
\label{eq-22}
{\tilde Q}_{ij}=\frac{1}{2} \left[
   \sum_{m=1-k}^{q} Z_{2q} (m) - \sum_{m=1+k}^p Z_{2p}(m)\right],
\end{equation}
where $k\equiv p-q=j-i>0$ and
\begin{equation}
\label{eq-23}
Z_{2n} (m)=\left( \begin{array}{c} 2n-m \\ n \end{array} \right)
           \left(\frac{1}{2}\right)^{2n-m},
\end{equation}
which  can be interpreted as a probability that a random
walker in one dimension
returns to the starting point
exactly $m$ times up to $2n$ steps \cite{Feller}.
Using identities (eqs.(A1) and (A2) in appendix), eq.(\ref{eq-22})
simplifies to
\begin{equation}
\label{eq-24}
{\tilde Q}_{ij}=\sum_{m=0}^{k-1}Z_{2p} (m)=1-\sum_{m=k}^p Z_{2p} (m).
\end{equation}

The diagonal elements ${\tilde Q}_{ii}$ can be written,
using eq.(\ref{eq-15}), as
\begin{eqnarray}
\label{eq-25}
{\tilde Q}_{ii} &=& 1+2\sum_{j=i}^{L-1} {\tilde Q}_{j,j+1} \nonumber\\
         &=& 1+2(p+1)Z_{2(p+1)} (0),
\end{eqnarray}
where eq.(\ref{eq-24}) and the identity of eq.(A3) are
utilized to derive the second equality. Eqs.(\ref{eq-24}) and
(\ref{eq-25})
form a complete set of exact solutions
for mass-mass correlation functions
in the steady state.

Now we are ready to calculate the fluctuation properties of
the surface in the RBD model. The surface height at site $i$,
$h_i (t)$,
can be written in terms of step-height variables $n_j(t)$ as

\begin{equation}
\label{eq-26}
  h_i(t)\equiv \sum_{j=1}^{i-1} n_j(t),
\end{equation}
where $i=2,\cdots,L$ and the reference height
$h_1(t)$ is set to be zero.
We consider two important fluctuations of the surface; the
step-height (mass) fluctuations and the height fluctuations.
\begin{eqnarray}
\label{eq-27}
M^2_i (L,t) &\equiv&
\langle n_i^2 \rangle_t - \langle n_i \rangle_t^2, \nonumber\\
W^2_i (L,t) &\equiv&
\langle h_{i+1}^2 \rangle_t - \langle h_{i+1} \rangle_t^2 \\
            &=& \sum_{j,j^\prime =1}^i \left[
           \langle n_j n_{j^\prime} \rangle_t -
\langle n_j \rangle_t \langle n_{j^\prime} \rangle_t \right].\nonumber
\end{eqnarray}
Notice that the fluctuations are site-dependent due to the lack of
the translational symmetry of our RBD model. So it is useful to
consider the fluctuations averaged over all sites.
The average mass fluctuation in the steady state is
\begin{equation}
\label{eq-28}
M^2(L,\infty) =\frac{1}{L-1} \sum_{i=1}^{L-1} M^2_i
= \frac{8}{3} L Z_{2L} (0),
\end{equation}
where eq.(\ref{eq-25}) and
the identity of eq.(A4) is used for the second equality.
For large $L$, $M^2$ grows like $L^{2\alpha^\prime}$ with
exponent $\alpha^\prime = 1/4$.
It is a little tricky to calculate the
average height fluctuation $W^2$ in the steady state.
However, using few identities of eqs.(A4)$-$(A7) and
eqs.(\ref{eq-24})
and (\ref{eq-25}), we find that
it is simply related to the average mass fluctuation as
\begin{equation}
\label{eq-29}
W^2(L,\infty)=2L-M^2(L,\infty).
\end{equation}
For large $L$, $W^2$ grows like $L^{2\alpha}$ with
exponent $\alpha=1/2$.

Time-dependence of the fluctuations, $M^2$ and $W^2$, can be
investigated numerically, using the time-dependent recurrence
relations in eq.(\ref{eq-12}). We take a surface with unit step
heights at every possible step
($n_i(0)=1$ for all $i$) as an initial configuration
which is equivalent to the ensemble-averaged surface
configuration in the steady-state limit
($\langle n_i \rangle_\infty=1$).
If some other surface configurations like a flat surface are
taken as an initial configuration,
the dynamics of the growing surface involves two different
time scales; one time scale associated with the evolution of
the ensemble-averaged surface into the steady-state configuration
and the other time scale associated with the development of
surface fluctuations. The former time scale is trivial and
uninteresting. Moreover, the scaling exponent $\beta$ in
eq.(\ref{eq-1}) characterizes the surface-fluctuation time scale.
We can single out the surface-fluctuation time scale by choosing
the above initial configuration. Starting with
$\langle n_i\rangle_0=1$ and $\langle n_i n_j\rangle_0=1$, we
solve the
time-dependent recurrence relations, eq.(\ref{eq-12}), iteratively.
In this paper, we consider the homogeneous case only; $\kappa_i=1/L$.
The results for the average mass fluctuation $M^2$ and the
average height fluctuation $W^2$ versus time $t$ are plotted
in Fig.4 (log-log scale) for system size $L=400$. We find
that $M^2\sim t^{2\beta^\prime}$ with
$\beta^\prime=0.253(5)\simeq 1/4$ and
$W^2\sim t^{2\beta}$ with $\beta=0.500(1)\simeq 1/2$.
In both cases, the dynamic scaling exponents,
$z=\alpha/\beta$ and $z^\prime=\alpha^\prime/\beta^\prime$,
are found to be
1  ($z=z^\prime=1$).

\section{Time scales in the RBD Model and the single-step model}

The RBD model can be viewed as a simple variant of the single-step
model
by rotating the surface by 45$^\circ$. As discussed in section II,
the RBD model with smooth particles instead of sticky particles
is equivalent to the single-step model except for boundary conditions.
Depositions occur only at the valleys of the 45$^\circ$-rotated
surface.
With $smooth$ particles (or in the single-step model), the height of
the valley increases by one unit at the deposition of a particle.
However,
with sticky particles in the RBD model, the heights of all sites
between the valley and the hill increase by one unit (see Fig.2).
In average, the deposition of a particle in the RBD model
is equivalent to the depositions of many particles
between the valley and the hill in the single-step model.
We argue that the number of depositions
in the single-step model corresponding to
a single deposition in the RBD model
is proportional to the average surface fluctuation width $W$.
In order to get the same morphology of the surface in average,
one should wait for much longer time $t_s$ in the single-step model
than in the RBD model. The elapsed time $t_s$ in the single-step model
is related to the elapsed time $t$ in the RBD model as
\begin{equation}
\label{eq-30}
t_s\sim  Wt.
\end{equation}
The surface fluctuation width of the
single-step model $W_s$ at time $t_s$ is equivalent to that of
the  RBD model at time $t$ in the infinite-size limit;
\begin{equation}
\label{eq-31}
W(\infty,t)=W_s(\infty,t_s).
\end{equation}
The single-step model belongs to the KPZ universality class with
scaling exponents $\alpha_{KPZ}=1/2$ and $\beta_{KPZ}=1/3$.
Using $W_s(\infty,t_s)\sim t_s^{\beta_{KPZ}}$, eq.(\ref{eq-31})
becomes
\begin{equation}
\label{eq-32}
W(\infty,t)\sim (Wt)^{\beta_{KPZ}},
\end{equation}
therefore
\begin{equation}
\label{eq-33}
W(\infty,t) \sim t^\beta,
\end{equation}
with
\begin{equation}
\label{eq-34}
\beta= \frac{\beta_{KPZ}}{1-\beta_{KPZ}}=\frac{1}{2}.
\end{equation}
It is shown that the scaling exponent $\beta$ is $1/2$
for the RBD model by simply comparing the deposition time scales in
the RBD model and the single-step model, which is consistent with
our result in the previous session. It implies that
these two models exhibit essentially the same scaling behavior
if the time-scale difference is taken into account. Therefore
the precise measurement of the exponent $\beta$ in the
RBD model (which is done in section IV) provides the good estimate
for the scaling exponents of the stochastic models in
the KPZ universality class.

We perform Monte Carlo simulations to confirm our argument
about the time-scale difference. We take the initial configuration
as a surface with $n_i(0)=1$ for all $i$ with horizontal length
$L=500$. After one deposition attempt on the average per lattice site
(one Monte Carlo Step), the time is incremented by one unit in the
time scale of the RBD model.
In order to use the time scale of the single-step model $t_s$,
the time increment for an actual deposition at site $i$
must be multiplied by
the step height $n_i (t_s)$. In this time scale, we run simulations
until $t_s=5,000$ and the number of independent runs is 3,000.
The double-logarithmic plot (Fig.5) for the
surface fluctuation width $W$
versus time $t_s$ shows a straight line with the slope
$\beta_s = 0.338(6)$, which agrees with the KPZ value $1/3$.
This result confirms our previous
argument about the time-scale difference between the RBD model and
the single-step model. Moreover, the mass fluctuation $M$ in this time
scale also shows the scaling behavior with exponent
$\beta_s^\prime=0.168(4)\simeq 1/6$. In both cases, the dynamic
exponents
$z_s$ and $z_s^\prime$ take the KPZ value $3/2$.
This result strongly supports
our suggestion that the RBD model exhibits essentially the same
scaling
behavior as the models in the KPZ universality class if the proper
time
scale is employed.

\section{Summary and Conclusion}

We have studied a restricted ballistic deposition (RBD) model
on a one-dimensional staircase with free boundary conditions.
Using generating function method,
the exact solutions for the surface fluctuation
and the mass fluctuation were obtained for any system size
in the steady-state limit. The RBD model is one of a few
nontrivial stochastic models which can be solved exactly.
We also examined time-dependent solutions
by solving the recurrence relations of two-point
correlation functions numerically.
{}From these solutions, we extracted the values of the scaling
exponents $\alpha$ and $\beta$, both of which
turned out to be $1/2$.

The RBD model is quite different from
the ordinary ballistic deposition (BD) model because the RBD model
does not allow the down steps.
Indeed, the scaling behavior of the
RBD model differs from the BD model
which belongs to the KPZ universality class where $\alpha_{KPZ}=1/2$
and $\beta_{KPZ}=1/3$. However, the RBD model can be viewed as
a simple variant of the single-step model which also belongs to
the KPZ universality class. These two models exhibit
a very similar surface morphology. And
their dynamical processes are not much different in average
except for the difference between deposition time scales.

We suggested that the major difference in the scaling behavior
of the RBD model and the single-step model may disappear
if the time-scale difference is properly taken into account.
So the static exponent
$\alpha$ should be equivalent in both models but the dynamic
exponent $\beta$ may be different.
We explored the
relation between two exponents as $\beta=
\beta_{KPZ}/(1-\beta_{KPZ})$, which is consistent with our results.
Monte Carlo simulations confirm our argument about the time-scale
difference.

The generalization of the RBD model into higher dimensions may
be quite interesting, partly because it may serve
as a useful alternative model for the indirect investigation of
the KPZ universality class in higher dimensions. Also our
generating function approach allows us to investigate the
models with inhomogeneous deposition rates.
Effect of the inhomogeneity on
the surface morphology and the scaling behavior in the RBD model is
currently under study.

\section*{Acknowledgments}

We wish to thank S. Redner for bringing this problem to our
attention. This work is supported in part by the BSRI,
Ministry of Education (No.94-2409), by the Korean
Science and Engineering Foundation (No.931-0200-019-2),
and by the Inha University research grant (1994).

\section*{Appendix}
Some useful identities are listed in this appendix.
$$
\sum_{m=0}^n Z_{2n}(m)=1.\eqno(A1)
$$
$$
\sum_{m=0}^{k-1} Z_{2n}(m)=\sum_{m=0}^{k-1}
Z_{2(n-k)}(-m-1).\eqno(A2)
$$
$$
\sum_{n=0}^{p-1} Z_{2n}(0)=2p Z_{2p} (0).\eqno(A3)
$$
$$
\sum_{n=0}^{p-1} nZ_{2n}(0)=\frac{2}{3} p(p-1) Z_{2p} (0).\eqno(A4)
$$
$$
\sum_{n=0}^{p-1} n^2 Z_{2n}(0)=\frac{2}{15} p(p-1)(3p-1)
Z_{2p} (0).\eqno(A5)
$$
$$
\sum_{m=0}^{n} mZ_{2n}(m)=(2n+1) Z_{2n} (0) -1.\eqno(A6)
$$
$$
\sum_{m=0}^{n} m^2 Z_{2n}(m)=-3(2n+1) Z_{2n} (0) +(2n+3).\eqno(A7)
$$

\newpage
\begin{center}{\large\bf Figure captions}
\end{center}

\begin{description}
\item[{\bf Fig.1 :}]
Dynamic processes of the RBD model. Depositions of particles
can occur only at the steps of the surface. The corresponding
particle dynamics model is also shown.

\item[{\bf Fig.2 :}] Deposition processes of
(a) the single-step model and
(b) the RBD model on the 45$^\circ$-rotated
surface.

\item[{\bf Fig.3 :}]
Recurrence relations are mapped to a directed random
walk problem in an upper diagonal half of
a $L\times L$ square lattice.
The thick line represents a typical path along which a directed
walker can start from $(L,L)$ and end at $(i,j)$.

\item[{\bf Fig.4 :}]  Log-log plots of the average mass
fluctuation $M^2(\bullet)$ and
the average height fluctuation $W^2(\circ)$ as functions of time $t$.
These data are obtained by solving the recurrence relations,
eq.(\ref{eq-12}).
The slopes of the lines are
$2\beta^\prime = 0.50$ and $2\beta = 1.00$.

\item[{\bf Fig.5 :}]  Log-log plots of the average mass
fluctuation $M^2(\bullet)$ and
the average height fluctuation $W^2(\circ)$
as functions of time $t_s$.
These data are obtained by Monte Carlo simulations using the
time scale of the single-step model.
The slopes of the best-fitted lines are
$2\beta^\prime = 0.336$ and $2\beta = 0.676$.

\end{description}

\end{document}